\documentclass[aps,twocolumn,nofootinbib]{revtex4}
\usepackage{amssymb}
\usepackage{amsmath}
\usepackage{graphicx}

\newcommand{\mean}[1]{\left\langle#1\right\rangle}

\newcommand{\dd}{\mathrm{d}}

\begin{document}

\title{Collision induced spatial organization of microtubules}
\author{Vladimir A. Baulin, Carlos M. Marques, Fabrice Thalmann}
\affiliation{Institut Charles Sadron CNRS UPR 22, 67083 Strasbourg
Cedex, France}
\date{\today}

\begin{abstract}
The dynamic behavior of microtubules in solution can be strongly modified by
interactions with walls or other structures. We examine here a microtubule
growth model where the increase in size of the plus-end is perturbed by
collisions with other microtubules. We show that such a simple mechanism of
constrained growth can induce ordered structures and patterns from an
initially isotropic and homogeneous suspension.  First, microtubules
self-organize locally in randomly oriented domains that grow and compete
with each other. By imposing even a weak orientation bias, external forces
like gravity or cellular boundaries may bias the domain distribution
eventually leading to a macroscopic sample orientation.
\end{abstract}
\maketitle

\section{Introduction}
\label{sec:introduction}

Biological processes like cell division, transport of certain organelles,
morphogenesis and organization in the cell are mediated by rod like structures
known as microtubules, which form various arrays, radial spindles, parallel
and antiparallel bundles \cite{book,Heald,Hyman}. The microtubule
self-assembly in living organisms is regulated by different factors:
microtubule-associated proteins (MAPs) which stabilize, destabilize and
crosslink microtubules \cite{Desai,JobRev}, diverse kinesin-like motor
proteins, which organize and link microtubules, $\gamma $-tubulin ring complex
which serves as a template for nucleation sites for microtubule polymerization
in centrosomes \cite{Canaday,Murata}. These factors combine with the physical
and chemical properties of the solution to determine, by mechanisms not yet
well understood, the spatial structure and the orientation of the
microtubules.

Each individual microtubule is a highly dynamic self-assembled
rod, which is permanently growing or shrinking. This ability for
being in an everlasting state of changing length as won
microtubules the name of "searching devices" for specific targets
in the cell \cite{Hyman,Vos}. A key property allowing for this
bistable state is the dynamic instability \cite{Desai}. Due to
conformational asymmetry of the constituting microtubule subunit,
the heterodimer $\alpha $,$\beta $-tubulin, a microtubule has a
polar structure and it grows at a \textit{plus}-end and shrinks at
its \textit{minus}-end. The speed of growth at the plus-end is not
constant, but rather intermittent. The elongation of the plus-end
is stochastically alternated by the abrupt shrinking, in a process
of unidimensional diffusion \cite{Maly}. Such dynamic behavior is
attributed to the complex, two-stage assembly of the plus-end,
implying the internal hydrolysis of the GTP in a tubulin dimer.
Within the cap model \cite{Desai,JobRev,Janosi}, the tubulins
added to the growing plus-end are not hydrolyzed, thus having
configurations favorable for the microtubule assembly. They form a
cap protecting the plus-end from the disassembly and shrinking.
Tubulins incorporated into microtubules are capable for
hydrolysis. During conversion of the GTP of $\beta $-subunit to
the GDP, the tubulin heterodimer undergoes the conformational
change that destabilize the tubular structure of a microtubule and
favors shrinking \cite{Krebs}.  The rate of hydrolysis compete
with the polymerization rate of the plus-end. Some factors like
local fluctuations of tubulin concentration near the plus-end may
perturb the balance thus leading to the loss of the protecting cap
and provoking rapid shrinking. Since the density fluctuations have
a stochastic nature, the growth to shrinking transitions
(catastrophes) and the inverse resume of growth (rescue) are also
statistically distributed.

The catastrophes can also be induced by factors other than
concentration fluctuations. During the growing state, microtubules
interact with each other as well as with different cellular
structures. The mechanical force opposing the growth of the
plus-end may induce catastrophes \cite{Dogterom,Tran}. Experiments
involving growing microtubules and different immobile obstacles
and barriers have shown that the opposing force reduces the growth
velocity \cite{DogtYrk,Dogt}.  Thus, the dynamic instability
coupled with the mechanical force may influence the microtubule
length and thus induce a spatial organization.  Indeed, the
catastrophe rate is expected to be higher near cellular boundaries
and lower in the cytoplasm \cite{Dogterom}. The boundaries may
also induce the orientation preference, since the catastrophe rate
in the direction perpendicular to the boundary would be higher
then that along the boundary. Recent in-vivo work suggests that
the microtubule dynamic instability is altered during preprophase
band formation \cite{Gadella}. Microtubule reorientation is
accompanied by the increase of the catastrophe frequency and
growth rate, while the rescue frequency and shrinkage rate remain
unchanged. The gradients always present in living cells can also
play a role of the "effective" boundaries and provoke the
microtubules ordering \cite{Karsenti}. In particular, the
gradients of the energy dissipation, concentration of tubulin and
associated proteins may result in spatial anisotropy of the
growing and shrinking speeds which can lead to a
self-organization.

Another important observation concerns the inter-microtubules interactions
\cite{Cyr}.  Encounters between cortical microtubules affect their dynamic
behavior. Steep contact angles of microtubules collisions provoke catastrophes
more often than the shallow contact angles, while the microtubules with close
angles have shown a tendency to zippering.

These observations suggest that the spatial self-organization of
microtubules might be induced by the coupling of the catastrophe
events and the configuration of neighboring microtubules.
Different examples of spontaneous self-organization of dynamic
microtubules have been reported in the literature
\cite{Job,Tabony,TabMag,Oostvelt,Cyr2}. One of them \cite{Tabony},
describes the formation of spatial structures in the in-vitro
solution of microtubules that start growing from seeds distributed
homogeneously. The resulting pattern strongly depends on the
direction of the gravitational field, where microtubules are
organized as highly aligned strips.  Interestingly, the observed
structures do not appear in weightlessness conditions. The authors
conclude from their observations that the Earth gravity only
triggers the symmetry breaking and does not affect individual
microtubules. The concentration of microtubules even coupled with
gravity is not sufficient to provoke the system orientation due to
excluded volume effects as in usual liquid crystals \cite
{Baulin,Baulin2}: the shaking or mixing of the sample irrevocably
destroys the pattern. Since the in-vitro preparation does not
contain any molecular motors or MAPs, the self-organization in
stripes is attributed to the dynamic nature of the microtubules:
the pattern formation disappears when the dynamic instability is
inhibited by the addition of taxol. Similar patterns are formed
under magnetic fields \cite{TabMag}. The in-vivo observations of
the reorientation of cortical microtubules in parallel arrays
\cite{Oostvelt,Cyr2,LloydGrav,Lloyd} are also ascribed to dynamic
instability: the depolymerization of disordered microtubules is
followed by the repolymerization into an ordered array. However,
the in-vivo self-organization can be regulated by MAPs or motor
proteins \cite{Murata,Wasteneys} or it may be a result of the
simultaneous action of many factors.

In this paper we explore the consequences of the simplest possible physical
hypothesis for explaining microtubule orientation from a coupling mechanism
between growth and inter-tubule interactions.  In the next section we set the
foundations for the physical model based on the observation that the
inter-microtubules collisions increase the rate of catastrophes~\cite{Cyr}. In
section~\ref{sec:results} we show numerically that this mechanism alone leads
to the orientation of microtubules in aligned stripes.  A theoretical
discussion of our main results is presented in section~\ref{sec:theory} and
our findings are summarized in the conclusion.

\section{The model}
\label{sec:model}

\subsection{A kinetically constrained growth model
\label{subsec:model}}

We propose a model based on the assumption that the assembly dynamics of a
particular microtubule is influenced by the configuration of the surrounding
microtubules. Our purpose is to exhibit the simplest possible model of
constrained growth inducing some self-organization of the microtubules. Thus,
in our model, we neglect all the other possible physical mechanisms that could
play a role in the co-alignment, such as collision induced turnover and
dislocation \cite{Oostvelt}, as well as the excluded volume interactions
leading to ordering in usual liquid crystals. The incorporation of these
factors would facilitate and enhance the alignment.

\onecolumngrid

\begin{figure}
\begin{center}
\resizebox{0.6\textwidth}{!}{\includegraphics{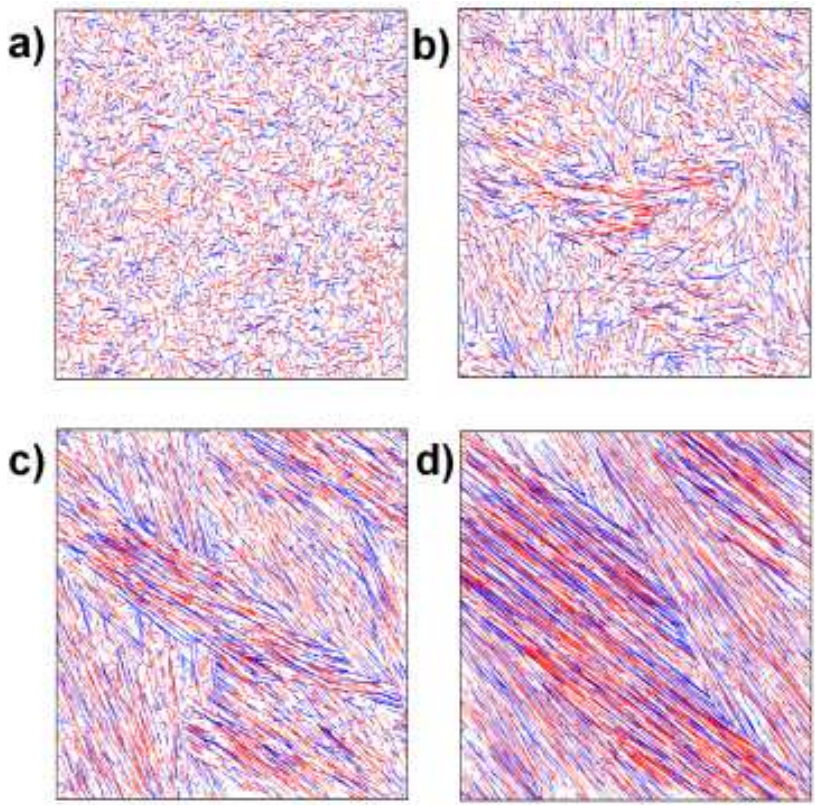}}
\caption{Snapshots of the system of rods at various successive
times (time increases from a to d). Growing rods are drawn in
blue, and shrinking rods in red.} \label{fig:snapshots}
\end{center}
\end{figure}

\twocolumngrid

A microtubule is modeled as a rigid, oriented rod which shrinks at
its minus-end and grows at its plus-end. Instead of dealing with a
fluctuating rate of growth and shrinking, we rather consider
smooth, coarse-grained properties, namely the average speeds of
growth and shrinking.  Without obstruction, in a free environment,
the plus-end of a microtubule grows at constant speed $v^{+}$,
while its minus-end shrinks at constant speed $v^{-}$.  When the
plus-end encounters another rod, it stops, but the rod continues
to shrink at its minus-end, with speed $v^{-}$, and as a result,
the overall length of the rod decreases. The rod resumes its
growth as soon as its plus-end is no longer blocked by its
neighbor. Altogether, the plus-end experiences an environment
dependent intermittent growth, and the minus-end a constant motion
at speed $v^{-}$. The rod disappears if its length decreases to
zero during the shrinkage phase. The total number of rods is not
fixed, but is maintained by a permanent injection rate of new rods
at random positions, with random orientation and zero length.

We found that this mechanism of the constrained growth can lead
itself to the spontaneous alignment of microtubules in an
initially isotropic and homogeneous array. The emergence of a
local orientational order is reminiscent from a natural selection
process. Whenever some local anisotropy builds up, the survival
rate of the neighboring rods changes and becomes orientation
dependent. The rods which from the start have picked up the
dominant orientation are likely to outlive rods with a different
orientation, and this creates conditions favoring the population
of rods with the ``correct'' orientation at the expense of rods
with ``incorrect'' orientation. In our model, the rods cannot
change their orientation but the permanent injection of young,
randomly oriented rods, leaves to the system the possibility to
reorganize and to tune up to a change of external conditions.

We implemented numerically this constrained growth model. The simulations
clearly show a trend towards some local organization and ordering, with the
formation of well defined anisotropic domains.

\subsection{Numerical implementation of the constrained growth model}
\label{subsec:numericalImplementation}

In our numerical implementation, in two dimensions, each rod is characterized
by its length, its position, and its orientation. The orientation (one angle)
and the position (two coordinates) are set when the rod is injected, and do
not change until the rod eventually disappears. The length of each rod evolves
with time, starting from zero shortly after the injection. All rods are packed
in a square box of side $L_s$, subject to periodic boundary conditions.

Depending on their dynamic state, the rods belong to two categories: shrinking
rods (\textit{s}-rods) are the rods that the local environment prevents from
growing further (kinetic constraint), while growing rods (\textit{g}-rods) are
free to grow.

The rule for the time evolution of the rods is the \textit{non-crossing
  displacement}.  Updates are done every time interval $\Delta t$. The minus
tip of each rod is shortened by the amount $v^{-}\Delta t$, while the plus tip
attempts a move forward by $v^{+}\Delta t$. If this move can be done without
crossing any other rod, the move is accepted and the rod grows. If the rod was
in a \textit{s}-rod state, it converts to a \textit{g}-rod state.  Otherwise,
the move is rejected, and the rod switches to, or stays in a blocked
\textit{s}-rod state.  As a result, the length of a rod after each step,
either increases by $(v^{+}-v^{-})\Delta t$, or decreases by $v^{-}\Delta t$.

In the present case, we chose the values $L_s=100$ and $\Delta
t=1$. The speed $v^+$ ranges from $0.5$ to $2.1$ and $v^-= 0.3$.
In the discussion, we make a frequent use of the speed ratio
$\alpha$:
\begin{equation}
\alpha = \frac{v^-}{v^{+}-v^{-}},
\label{eq:definitionAlpha}
\end{equation}
defined as the ratio between the speed of shrinkage in the \textit{s}-state
$v^-$ and the speed of growth in the \textit{g}-state $v^+-v^-$. The
corresponding values of $\alpha$ used in the simulation lie in the interval
0.17 to 1.5.  The value for the injection rate $Q_i$ is about 1000 new rods
per unit of time. There are typically a few thousand rods (5000 to 6000) at a
time in the simulation box.

\section{Main results}
\label{sec:results}

\subsection{Spatial organization: domain structure}

The simulation starts with a set of rods of zero length. The kinetic
constraint concerns a vanishingly small number of rods at the early stage of
the system evolution. As the number and the length of the rods increase, the
amount of packing gets larger, and the kinetic constraint forces a significant
fraction of rods into a blocked, shrinking state. This transient regime
recedes to a quasi-stationary regime in which the ratio of \textit{s}-rods and
\textit{g}-rods seems to remain approximately constant.

Then, the numerical simulations shows clearly the slow emergence of oriented
domains, or bundles, of nearly aligned rods. These domains show very rough,
ragged and sharp boundaries, much as crystallites. Thus, they look quite
different from the domains arising in the usual phase transitions and
coarsening situations, where a finite bending elasticity creates a smooth
variation of the order parameter, at the vicinity of a domain wall.  The
domains seems to be randomly oriented, and the isotropy of the system is
recovered only on length scales larger than the size of the domains.

In the final stages of the simulation, the average domain size $L(t)$ is still
a slowly growing function of the time, and eventually becomes of the same
order of magnitude as the size of the system $L_s$. This prevents reaching an
asymptotic finite value of $L(t)$, associated to a truly stationary
distribution of the rod lengths and orientations. Well known examples of such
coarsening dynamics are characterized, for instance, by a power-law or a
logarithmic behavior of $L$ with $t$~\cite{Bray}. In the latter case, the
possibility to discuss the system properties in term of quasi-stationary
solutions remains.

\begin{figure}
\begin{center}
\resizebox{7cm}{!}{\includegraphics{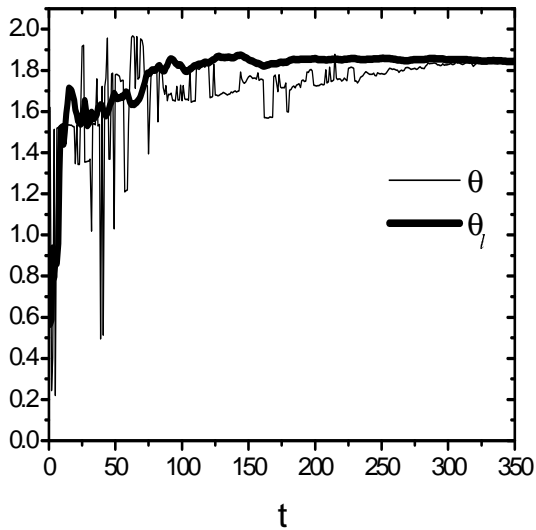}} \caption{The
dominant angles $\Theta$, $\Theta_l$ as a function of the time, as
found by minimizing equations~\protect\ref{eq:minimizationTheta}
  and~\protect\ref{eq:minimizationThetaL}.}
\label{fig:curvesTheta}
\end{center}
\end{figure}

Figure~\ref{fig:snapshots} illustrates how the rods self-organize
with time. At first, the population of rods is isotropic, except
for small fluctuations inherent to the random initial position and
orientation distribution (Figure~\ref{fig:snapshots}a). These
pre-existing heterogeneities grow into small bundles, whose
distribution still remain seemingly isotropic on large scales
(Figure~\ref{fig:snapshots}b). Then, larger bundles emerge at the
expense of many other smaller bundles, bound to disappear
(Figure~\ref{fig:snapshots}c). Finally, the typical size of the
larger bundles becomes comparable to the size of the simulation
box (Figure~\ref{fig:snapshots}d). The absence of bending modulus,
and the presence of ragged boundaries, forbids a mechanism based
upon domain walls motion. Such a competitive growth of the domains
is slow.

\begin{figure}
\begin{center}
\resizebox{7cm}{!}{\includegraphics{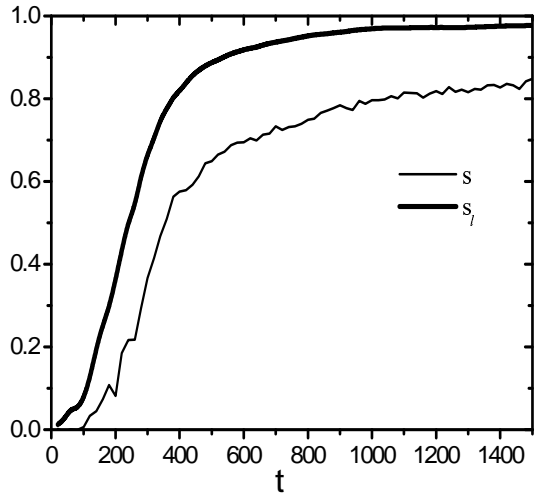}} \caption{Anisotropy
ratios $S_l$ and $S$ \textit{vs} time $t$.}
\label{fig:curvesAnisotropyRatio}
\end{center}
\end{figure}

To quantify the degree of local ordering, or ``polarization'', of the system,
we introduce a dominant angle $\Theta$, which maximizes a cost function
$\sigma$:
\begin{eqnarray}
\sigma (\theta )&=&
\frac{1}{n_t}\sum_{i=1}^{n_t}\cos^{2}\left(\Omega_{i}
  -\theta\right),\nonumber\\
&=& \overline{ \cos^{2}\left(\Omega  -\theta\right)},
\end{eqnarray}
where $\Omega_i$ is the orientation (angle) of the $i^{th}$ rod, and the sum
runs over $n_t$, the total number of rods present in the system. Thus,
$\sigma$ is defined as the ``ensemble average'' over the population of rods at
time $t$, denoted with an overline $\overline{\vphantom{l}\ldots}\;$ We call
order parameter the value of the maximum $s=\sigma(\Theta)$. The cost function
can be expanded as $\sigma(\theta)=\overline{\cos^{2}\Omega} \cos^{2}\theta
+\overline{\sin^{2}\Omega} \sin^{2}\theta + \overline{2\sin
  \Omega\cos\Omega}\sin^2\theta$.  Then, the stationarity condition
$\partial\sigma(\theta)/\partial\theta|_{\Theta}=0$ leads to:
\begin{equation}
\tan 2\Theta =\frac{\overline{\sin 2\Omega}}{\overline{\cos 2\Omega}}.
\label{eq:minimizationTheta}
\end{equation}
This equation has always four solutions, two corresponding to the maxima
$\Theta_{\max }$ and $\Theta_{\max }+\pi$, and the other two, to the minima
$\Theta _{\min }$ or $\Theta _{\min }+\pi$, and $\Theta_{max}$ and
$\Theta_{min}$ are mutually orthogonal.

A scaled anisotropy parameter $S$ may be defined as:
\begin{equation}
S =\frac{\sigma (\Theta _{\max })-\sigma (\Theta _{\min })}{\sigma (\Theta
_{\max })+\sigma (\Theta _{\min })}.
\label{eq:anisotropyRatio}
\end{equation}
The quantity $S$ is~0 for a population of isotropically oriented rods, while
it is~1 for a population of perfectly aligned rods. The parameter $S$ makes it
possible to quantitatively assess the amount of ordering in the system.
Because both parameters $s$ and $S$ turn out to fluctuate strongly with time,
we introduce  a more stable parameter, where each rod $i$ contributes
according to its length $l_i$:
\begin{equation}
\sigma_l(\theta)=  \frac{1}{n_t}\sum_{i=1}^{n_t}l_{i}^{2}\cos^{2}\left( \Omega
  _{i}-\theta \right),
\label{eq:orderParameterL}
\end{equation}%
where the long rods participate more than the short ones.
The dominant angle associated with this parameter obeys:
\begin{equation}
\tan 2\Theta _{l}=\frac{\overline{l^{2}\sin 2\Omega}}
{\overline{l^{2}\cos 2\Omega}},
\label{eq:minimizationThetaL}
\end{equation}
and the anisotropy ratio $S_l$ defined as in
Eq.~\ref{eq:anisotropyRatio}, with $\sigma$ replaced by
$\sigma_l$.

Figure~\ref{fig:curvesTheta} shows the variation of the dominant angles
$\Theta$ and $\Theta_{l}$ with time, for one set of parameters. At the
beginning, the system is homogeneous and the distribution of angles is
isotropic, resulting in a singular and noisy function of time.  As the system
evolves, the ordered structures appear and the angles stabilize around their
preferred value. It is noteworthy that the $\Theta _{l}$ curve is smoother
than the $\Theta$ curve, due the stabilizing contribution of the longest and
most stable rods. The plateau value is related to the orientation of the
dominant bundle, and fluctuates from sample to sample.

The same conclusion can be drawn from the plot of the anisotropy
ratios $S$ and $S_l$, function of time in
Figure~\ref{fig:curvesAnisotropyRatio}. Although evolving on the
same time scale as $S$, the quantity $S_l$ reaches a value closer
to~1. The differences between the two curves is most certainly due
to the contribution of the many young, short rods, upon which the
kinetic constraint has not been acting long enough to force them
into the dominant orientation.

\begin{figure}
\begin{center}
\resizebox{7cm}{!}{\includegraphics{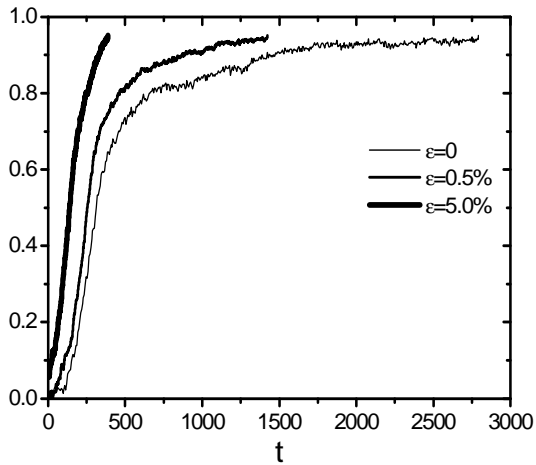}} \caption{The
anisotropy ratio $S_l$ \textit{vs} time, for three increasing
biases $\epsilon$ in the angular distribution of the rods
orientation.} \label{fig:effectBias}
\end{center}
\end{figure}

The anisotropy ratio and the dominant angles aim at quantifying the degree of
local ordering in a suspension of rod like objects, irrespective of the
underlying alignment mechanism. In that respect, they make possible a direct
comparison with other alternative models of microtubule orientation. From
these curves, one can infer a characteristic time $t^*$ for the emergence of a
global orientation in the sample, such as, for instance, $S_l(t^*)=1/2$. In
Figure~\ref{fig:curvesAnisotropyRatio}, this ordering time is about a few
hundred steps ($t^*\sim 300$).

\subsection{Sensitivity to external stresses}
\label{subsec:sensitivity}

The constant renewal of the rods, along with the growth of the competing
domains, confers to the system the ability to respond to external
perturbations. One of our main motivation is to evaluate the sensitivity of
the ordering to the presence of an external gravitational, or magnetic field.
Quite similarly, the presence of a hard wall is expected to align the
nearby domains along its direction.

\begin{figure}
\begin{center}
\resizebox{0.4\textwidth}{!}{\includegraphics{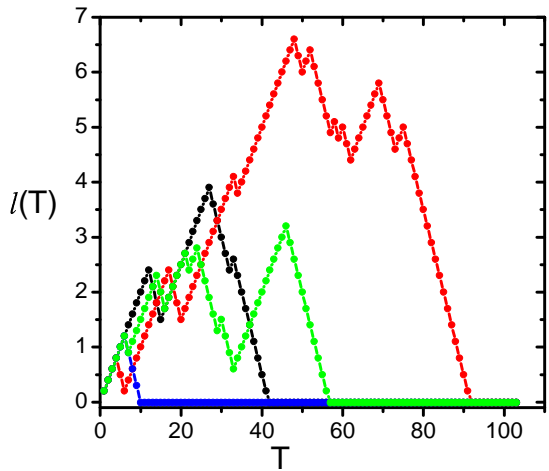}}
\caption{Different rod histories, showing the length $l(t)$
function of the time $t$. The curves are similar to random walks
with an absorbing boundary condition at $l=0$.}
\label{fig:lifeRod}
\end{center}
\end{figure}

In order to probe the ability of the rods suspension to cope with
external constraints, and to monitor its susceptibility to a small
symmetry breaking, we performed several simulations with a
slightly biased distribution in the orientation of the newly
injected rods. Instead of being isotropic, we added a fraction
$\epsilon$ of rods in excess, with an angle $\Omega$ belonging to
a small interval $\Omega_0\pm 1^{\circ}$.  As a result, a value
$\epsilon=0.5\%$ brings about an acceleration of the ordering time
$t^*$ by a factor $1.5$, and a value $\epsilon=5\%$ triggers a
three times faster growth of the domains
(Figure~\ref{fig:effectBias}). In both cases, the preferred
orientation is clearly related to the orientation of the  bias
$\Omega_0$.

\begin{figure}
\begin{center}
\resizebox{0.45\textwidth}{!}{\includegraphics{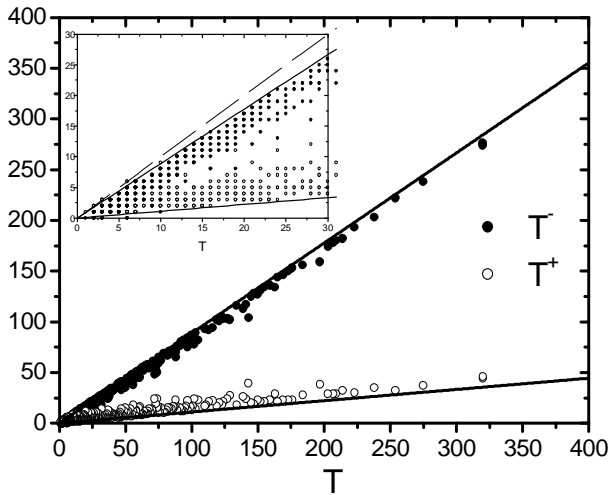}}
\caption{Scatter plot of the times of growth $T^+$ and $T^-$, as a
function of the total age $T=T^{+}+T^{-}$, for a given population
at time $t$. Inset: a close-up look at the region of young rods.}
\label{fig:scatterAges}
\end{center}
\end{figure}

Boundaries and impurities can also induce the alignment. When one of the
periodic boundary conditions is replaced by a hard wall, the alignment of the
rods is much faster, and the wall orientation propagates into the bulk of the
suspension. An identical behavior is observed when a rod with fixed length,
position and orientation, is forced into the simulation box. The rods orient
themselves parallel to the guiding rod, and longer guiding rods provoke faster
ordering.

\subsection{Kinetics of individual rods}

The numerical simulation makes it possible to track a single rod
as it evolves with time. We observe that during its life cycle, a
rod can experience many alternating periods of growth and
shrinkage. A typical microtubule life history plot is shown in
Figure~\ref{fig:lifeRod}, following a saw-teeth curve.

\begin{figure}
\begin{center}
\resizebox{0.45\textwidth}{!}{\includegraphics{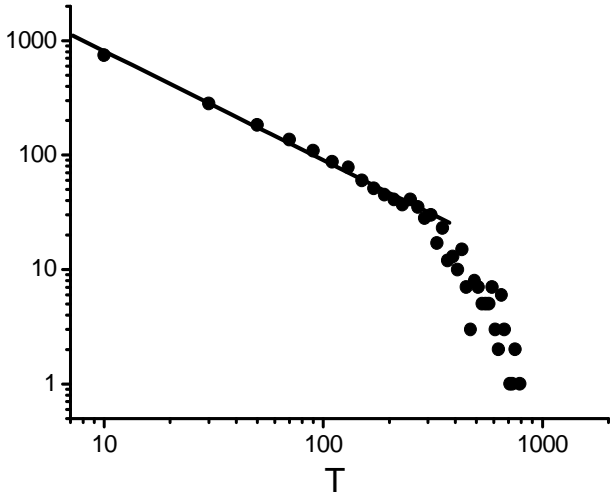}}
\caption{Histogram of the distribution of the ages $T$ of a
population of rods, in logarithmic coordinates. The initial decay
rate is close to $1/T$, followed by an exponential decay.}
\label{fig:timeDistribution}
\end{center}
\end{figure}

\begin{figure}
\begin{center}
\resizebox{8cm}{!}{\includegraphics{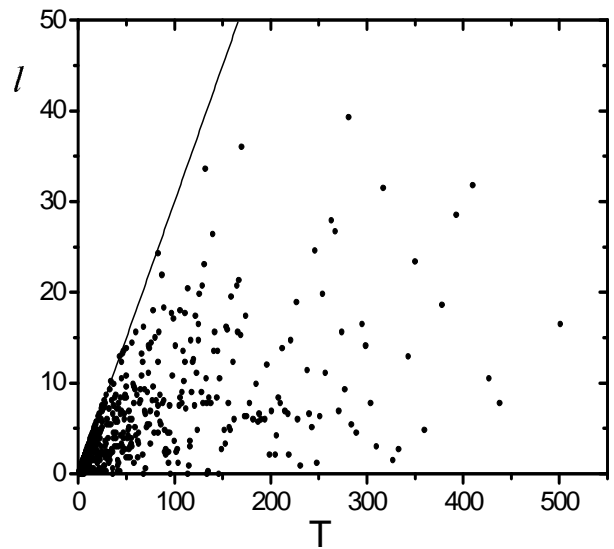}} \caption{Scatter
plot of the length (vertical axis) and the age (horizontal axis).}
\label{fig:scatterLengths}
\end{center}
\end{figure}

We call ``age'' $T$, the time interval elapsed since the rod was injected with
zero length in the system. The age is the sum of the growing time $T^+$ and
the shrinking time $T^-$, and the length of the rods can be expressed with the
help of $T^+$, $T^-$ as
\begin{eqnarray}
  l&=&(v^{+}-v^{-})T^{+}-v^{-}T^{-};\nonumber\\
  T&=&T^{+}+T^{-},  \label{eq:lT}
\end{eqnarray}
or equivalently,
\begin{equation}
T^{-} = \frac{(v^{+}-v^{-})T-l}{v^{+}};\;
T^{+} = \frac{v^{-}T +l}{v^{+}}.
\label{eq:TpTm}
\end{equation}

A typical distribution of both $T^{+}$ and $T^{-}$, as a function
of the age $T$, is shown in Figure~\ref{fig:scatterAges} for a
population of rods at a given time $t$ (scatter plot), while the
inset of Figure~\ref{fig:scatterAges} is an enlargement of this
plot in the small $T$ region. The values $T^{+}$ and $T^{-}$ of
old rods (large $T$), concentrate near two boundaries, which
correspond to a length $l=0$ in Eq.~\ref{eq:TpTm}. For these old
rods, which have survived many collisions, the growth and
shrinkage periods compensate almost exactly, and in
Eq.~\ref{eq:lT}, the length $l$ results from the difference
between two large quantities.  By contrast, the young rods (small
$T$) show all possible combinations of $T^{+}$ and $T^{-}$ (inset
in Figure~\ref{fig:scatterAges}).  This indicates that the young
rods have not yet been influenced by their surrounding. The
maximal possible length of a rod occurs in the extreme case
$T^{-}=0$ and $T^{+}=T$, thus corresponding to a length
$l_{\max}=(v^{+}-v^{-})T$ (the dashed line in the inset of
Figure~\ref{fig:scatterAges}).

Young rods enjoy a fast growth rate, but many are also eliminated
quickly. Older rods show a smaller average growth rate, but their
survival rate increases with their age. This is clear from the
histogram of the ages, which is clearly not exponential, but
rather well approximated by a power law $p_s(T) \sim T^{-1}$
(Figure~\ref{fig:timeDistribution}). The relative disappearance
rate of rods aged $T$, is $p^{-1}_s \dd p_s/\dd t \sim T^{-1}$,
instead of remaining constant, as in the exponential case
(\textit{e.g} like for instance the decay of radioactive
elements).

\begin{figure}
\begin{center}
\resizebox{7.2cm}{!}{\includegraphics{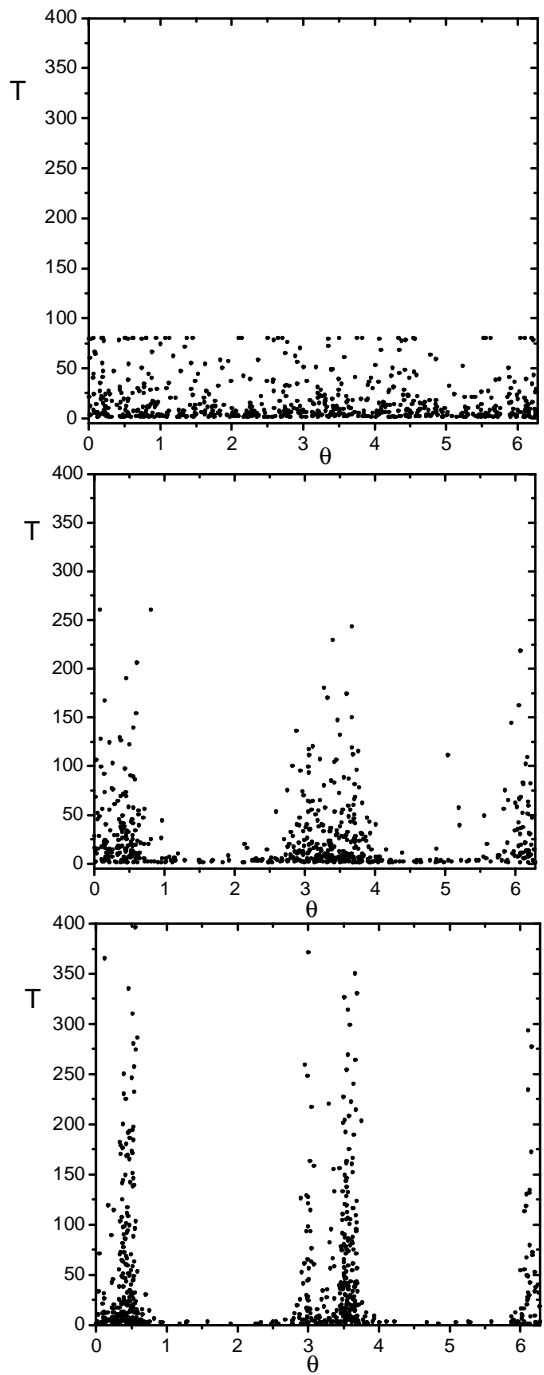}} \caption{Scatter
plot of the orientation (horizontal axis) and the age (vertical
axis) of the rods at three different stages of the evolution of
the system. The time $t$ increases from top to bottom.}
\label{fig:scatterAnglesAges}
\end{center}
\end{figure}

In a sense, the young rods shows ``plastic'' properties, and
account for the adaptability properties of the rods suspension. By
contrast, older rods are expected to show more rigidity and
persistence from the past history of the suspension. We did not
find any simple justification for the exponent $-1$, which is a
numerical finding. This contradicts a naive argument based on the
usual random walks, which would predict a $T^{-3/2}$ behavior
associated to the first return to the origin time distribution. We
still believe in the analogy, but we attribute this discrepancy to
the existence of strong correlations between orientation, age and
life time of the rods. Finally, the age of the very old rods is
distributed exponentially (Figure~\ref{fig:timeDistribution}).

A scatter plot of the lengths~$l$ \textit{versus} the ages~$T$ of
a population of rods does not support the presence of strong
correlations between these two parameters
(Figure~\ref{fig:scatterLengths}).

\subsection{Distribution of lengths and orientations}

The ordered structures (bundles or domains) effectively select the rods, keeping only those
with an orientation compatible with the dominant orientation of the bundles.
The correctly oriented rods collide less often with their neighbors than the
rods with transverse orientations, and their ``fitness'', or survival ability
is greater.

Figure~\ref{fig:scatterAnglesAges} shows three typical scatter plots of the
ages as a function of the angles. The age of the system increases from the top
to the bottom plot. The presence of anisotropic domains manifests itself as
sharp peaks around a few well defined angles. On the example shown, one can
see two different domains, respectively around $25^{\circ}$ and $160^{\circ}$.
The peaks are duplicated (mirrored) because the bundles contain a mixture of
two antiparallel populations, separated by $180^{\circ}$.

\begin{figure}
\begin{center}
\resizebox{0.43\textwidth}{!}{\includegraphics{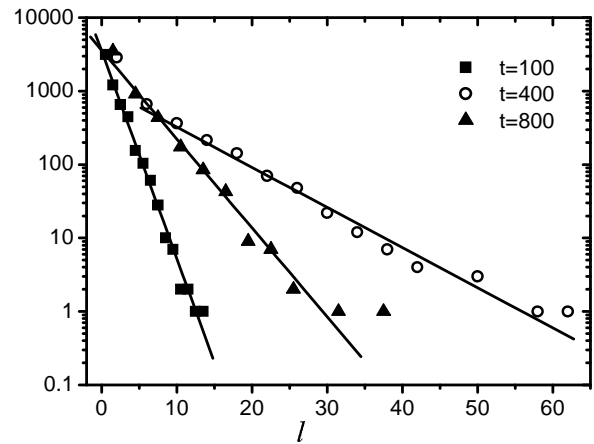}}
\caption{Semilogarithmic plot of the histogram of the number of
rods, for three different stages of the evolution of the system.
Straight lines correspond to an exponential decay.}
 \label{fig:histogramLengths}
\end{center}
\end{figure}

A typical distribution of lengths regardless to the rods
orientation is shown in Figure~\ref{fig:histogramLengths}. The
distribution is exponential except for a very small region of tiny
lengths. Most of the rods with tiny lengths are very young rods
injected in the system a few steps ago. They did not have time to
experience collisions and their distribution did not acquire the
same characteristics as the old rods.

\begin{figure}
\begin{center}
\resizebox{0.45\textwidth}{!}{\includegraphics{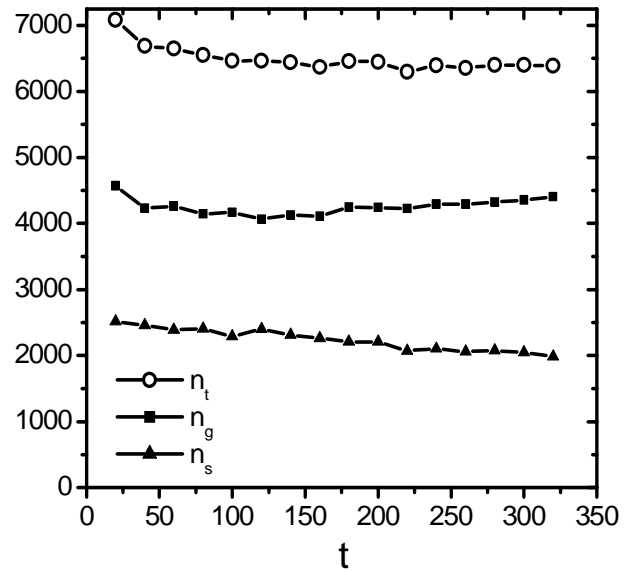}}
\caption{Evolution of the number of rods $n_t$ with time, and
repartition between shrinking rods ($n_s$) and growing rods
($n_g$).} \label{fig:numberRods}
\end{center}
\end{figure}

The system not only adjusts the orientation and the length of the
rods, but it also regulates the total number of rods $n_t$. An
increase in the rods density leads to a higher collision rate
among rods, and a higher elimination rate. In the early stage of
the simulation, in a sparse system, the newly injected rods do not
meet any obstacles and the total number of rods increases sharply.
After a transient regime, $n_t$ evolves very slowly, although it
is not strictly constant. As a matter of fact, $n_t$ slightly
decreases with the spreading of the dominant orientation and the
emergence of domains (Figure~\ref{fig:numberRods}).

\subsection{The three dimensional case}

Finally, we performed a limited number of simulations in three
dimensions, in order to check whether the kinetic ordering was a
special feature of the two dimensional systems, or whether it was
a generic feature also in three dimensions. In this case, our
simulation box is a cube of size $100\times 100\times 100$. In
addition to the injection rate $Q_i$, to the speeds $v^+$ and
$v^-$, there is another relevant parameter: the diameter, or
thickness, of the rods $d$.

It turns out that the behavior of the system in three dimensions is quite
similar to the one observed in two dimensions. The initially homogenous
solution becomes gradually structured into bundles and domains. However, the
three dimensional system differs by the absence of sharp boundaries between
domains. The competition between the different orientations is not so drastic,
since bundles with different orientation can interpenetrate if the rod
thickness is small. Domain walls are more difficult to identify, but the main
result, \textit{i.e.} local ordering, holds also in three dimensions.

\section{Theoretical discussion}
\label{sec:theory}

We discuss in this part some observed features of our numerical simulations:
exponential tails in the length distribution, collision rates, anisotropy. For
this purpose, we propose an elementary kinetic theory, and its predictions are
compared with the numerical simulations.

\subsection{Ensemble and time averages}
Much like in the usual statistical mechanics, one introduces two kinds of
averages. The time average, denoted by brackets $\langle\ldots\rangle$
corresponds to the mean value obtained by the repeated observation of single
rods evolving with time. For instance, $\langle T^{+}\rangle$ is the average
growth time of a rod. Time averages are accessible through numerical
simulations.

The ensemble average amounts to considering the whole population of rods at a
given time. This corresponds to an instantaneous ``snapshot'' of the
population of rods, and the corresponding average is denoted with an overline
$\overline{\vphantom{l}\ldots\;}$: for instance, $\overline{l}$ is the average
length of the rods. In practice, the ensemble average consists in summing over
all the rods, and then dividing by their total number $n_t$. The ensemble
averages can be computed from the simulations, but are also the natural
outputs of the kinetic theory sketched below.

The connection between time and ensemble average is by no means obvious. In
the case of a true stationary situation, both averages are expected to
coincide.  Our situation, however is not a full stationary situation, as we witness
the emergence of unbounded large bundles. A slow coarsening dynamics, however,
may still exhibit a satisfactory agreement between the two kinds of averages.

Ensemble averages are conveniently handled by means of distribution
functions. Denoting the length, orientation and position of the rods
respectively by $l$, $\Omega$ and $\vec{r}$, we define $c_t(l,\Omega
,\vec{r},t)= c_{g}(l,\Omega,\vec{r},t)+c_{s}(l,\Omega ,\vec{r},t)$, where
$c_g$, $c_s$ and $c_t$ stand respectively for the distribution of the
population of growing rods, shrinking rods and total number of rods, per unit
of surface, at time $t$,  with $ 0\leq l<\infty$, $0\leq\Omega<2\pi$ and
position $\vec{r}$.

Successive integrations over the variables $\vec{r}$, $l$ or $\Omega$, give
rise to a hierarchy of distribution functions. In particular, the numbers of
rods $n_s$ and $n_g$ are given by:
\begin{eqnarray}
n_{g}(t) &=&\int \dd\vec{r}\ \dd l\ \dd\Omega\, c_{g}(l,\Omega ,%
\vec{r},t);  \label{n} \\
n_{s}(t) &=&\int \dd\vec{r}\ \dd l\ \dd\Omega\, c_{s}(l,\Omega ,%
\vec{r},t).
\end{eqnarray}

In what follows, we use a loose notation for the partial distribution
functions, where the variables which do not explicitly appear in $c_g$ have
been implicitly integrated over, \textit{e.g.} $c_g(l,\Omega)\dd l\ \dd
\Omega$ stands for the fraction of \textit{g}-rods, with length between $l$
and $l+\dd l$, angle between $\Omega$ and $\Omega+\dd \Omega$, but located at
any position $\vec{r}$ of the system.

\begin{figure}
\begin{center}
\resizebox{0.45\textwidth}{!}{\includegraphics{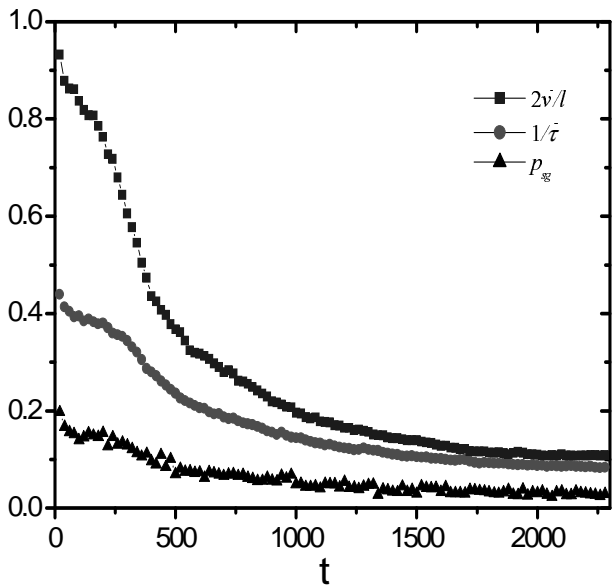}}
\caption{Test of the relations Eq. \protect\ref{eq:psg:intro} and
Eq. \protect\ref{eq:psg}. Circles: inverse of the average
shrinkage interval $\langle\tau^-\rangle$, last term of Eq.
\protect\ref{eq:tauRate}, squares: r.h.s of Eq.
\protect\ref{eq:psg:intro}, triangles: computed value of
$p_{sg}$.} \label{fig:testpsg}
\end{center}
\end{figure}

\subsection{The distribution of lengths}
\label{subsec:expo-length}

Following the lines of Appendix~\ref{app:kineticTheory}, we restrict ourselves
to the homogeneous, $\vec{r}$-independent case, and assume that the
distributions $c_g$ and $c_s$ comply with the following master equation:

\begin{equation}
\left\{
\begin{array}{c}
\frac{\partial}{\partial t}c_{g}  +v\frac{\partial}{\partial l}c_{g}%
=p_{sg}c_{s}-p_{gs}c_{g};\\
\frac{\partial}{\partial t}c_{s}-\alpha v\frac{\partial}{\partial l}c_{s}%
=p_{gs}c_{g}-p_{sg}c_{s}.
\end{array}
\right. \label{eq:masterEquation}
\end{equation}

The properties of the distributions mostly depend on the injection
rate $Q_i$, the speed $v=v^+-v^-$, and the speed ratio
$\alpha=v^{-}/(v^+-v^-)$. We introduce the interconversion rates
$p_{sg}(\Omega)$ and $p_{gs}(\Omega)$ between \textit{s} and
\textit{g}-states, and we take care of a possible dependence on
the direction $\Omega$.  We find that, for an homogeneous and
stationary system, the lengths are exponentially distributed and
verify:
\begin{eqnarray}
c_g &=& \frac{Q_i}{v} \exp[-l/\overline{l}(\Omega)];\\
c_s &=& \frac{Q_i}{\alpha v} \exp[-l/\overline{l}(\Omega)].
\end{eqnarray}

Such an exponential distribution can be seen in
Figure~\ref{fig:histogramLengths}. This model accounts for the
possibility of anisotropic length distributions, via the angle
dependent function $\overline{l}(\Omega)$. It is possible to find
an isotropic, self-consistent solution for $\overline{l}(\Omega)=
\overline{l}$, but we found also evidence for an anisotropic
self-consistent solution, with a non trivial function
$\overline{l}(\Omega)$.

\begin{figure}
\begin{center}
  \resizebox{0.45\textwidth}{!}{\includegraphics{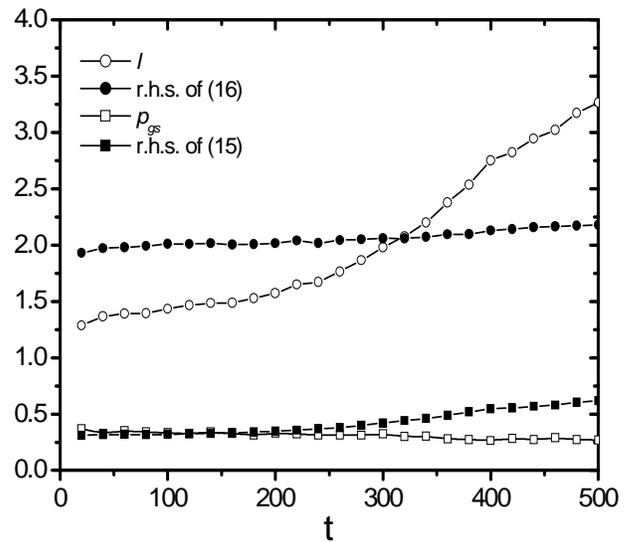}}
\caption{Test of the relations Eq.
\protect\ref{eq:lp-isotropic:intro} and Eq.
  \protect\ref{eq:l-iso-selfconsistent}.
  Open circles: $\overline{l}$, filled circles: r.h.s of Eq. \protect\ref{eq:l-iso-selfconsistent},
  open squares: $p_{gs}$,
filled squares: r.h.s of Eq. \protect\ref{eq:lp-isotropic:intro}.}
\label{fig:testpgs}
\end{center}
\end{figure}

Our predictions for the isotropic solution include a determination of the
rate $p_{sg}$:
\begin{equation} p_{sg} =\frac{2v^{-}}{\overline{l}},
\label{eq:psg:intro}\end{equation}
a determination of the rate $p_{gs}$:
\begin{equation}
p_{gs}(\Omega )=v^{+}\frac{2\overline{l}}{\pi}
\left(\frac{n_t}{S}\right),
\label{eq:lp-isotropic:intro}
\end{equation}
and a self-consistent determination of the average length $\overline{l}$, as a
function of $Q_i$, $\alpha$, the total number of rods $n_t$ and the surface
$S$:
\begin{equation}
\overline{l}=\sqrt{\frac{3\pi}{2(\alpha+1)}\frac{S}{n_t}} \simeq
\left(\frac{S}{n_t}\right)^{\frac{1}{2}}.
\label{eq:l-iso-selfconsistent}
\end{equation}

The predictions of Eq.~\ref{eq:psg:intro} are shown in
Figure~\ref{fig:testpsg} and discussed also in
Appendix~\ref{app:stationaryCase}.  Agreement is poor for short
times and it improves for long times.

The predictions of Eq.~\ref{eq:lp-isotropic:intro} and
Eq.~\ref{eq:l-iso-selfconsistent} are summarized in
Figure~\ref{fig:testpgs}. The agreement is good for $p_{gs}$ and
qualitative for $\overline{l}$ at short times. The prediction
becomes poor for times larger than $t^*$, associated to the
emergence of the domains. We believe that the disagreement is
mainly due to the impossibility for a two-dimensional system to be
at the same time considered as anisotropic and homogeneous. By
contrast, this would be a more reasonable assumption in a three
dimensional space. Our kinetic theory shows too strong mean-field
features to be able to describe accurately this two-dimensional
situation. We conclude that the predictions of this isotropic
model are no longer valid when the domains start growing.

\subsection{Connections with the individual history of the rods}


\begin{figure}
\begin{center}
\resizebox{0.45\textwidth}{!}{\includegraphics{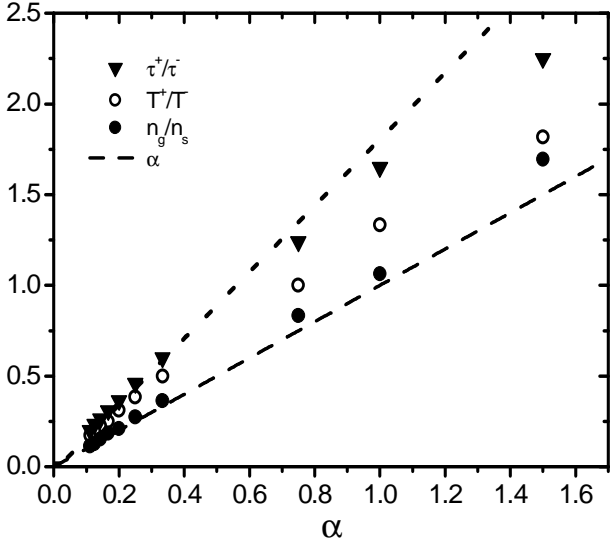}}
\caption{Horizontal axis: speed ratio $\alpha=v^-/(v^+-v^-)$.
Various curves are represented: the ratio of the times of growth
to the times of shrinking $\mean{T^+}/\mean{T^-}$, the ratio of
intervals of growth and the intervals of shrinking
$\mean{\tau^+}/\mean{\tau^-}$, the ratio of the number of growing
and shrinking rods $n_g/n_s$ and $\alpha$ itself. This Figure
illustrates Eq.~\protect\ref{eq:relationNgNsTaupTaumTpTm}.}
\label{fig:populations}
\end{center}
\end{figure}

On the graph showing the individual history of the rods
(Figure~\ref{fig:lifeRod}), one can decompose the time of growth
$T^+$ into a sum of elementary growth intervals $\tau_j^+$, and
the time of shrinking $T^-$ into a sum of elementary shrinking
intervals $\tau_j^-$:
\begin{equation}
T^+ = \sum_j \tau_j^+;\;
T^- = \sum_j \tau_j^-.
\label{eq:decompostionTpTm}
\end{equation}

In particular, we expect that the following relations between the average
elementary time of growth $\langle \tau^+\rangle$, shrinkage $\langle
\tau^-\rangle$, and the rates $p_{gs}$ and $p_{sg}$ hold for an isotropic
system:
\begin{equation}
p_{gs} = \langle\tau^+\rangle^{-1};\; p_{sg} = \langle\tau^-\rangle^{-1}.
\label{eq:tauRate}
\end{equation}

For an anisotropic system, $p_{gs}$ becomes orientation dependent, while
$p_{sg}$ should not. In a stationary case, the following ``detailed balance''
relation holds:
\begin{equation}
n_g p_{gs} = n_s p_{sg},\label{eq:detailedBalance}
\end{equation}
while in a quasi-stationary state, we expect only a qualitative relation:
\begin{equation}
\frac{n_g}{n_s} \simeq \frac{
  p_{sg}}{p_{gs}},\label{eq:detailedBalance:quasistat}
\end{equation}
suggesting the relation:
\begin{equation}
\frac{n_g}{n_s} \simeq \frac{ p_{sg}}{p_{gs}}=
\frac{\langle\tau^+\rangle}{\langle\tau^-\rangle}.
\label{eq:relationNgNsTaupTaum}
\end{equation}

In any case, from Eq.~\ref{eq:lT}, we have:
\begin{eqnarray}
\left\langle l\right\rangle &=&  (v^{+}-v^{-}) \left\langle T^-\right\rangle +
v^{-}\left\langle T^-\right\rangle,\label{ll} \\
\left\langle T\right\rangle &=& \left\langle T^+\right\rangle + \left\langle
T^-\right\rangle.\nonumber
\end{eqnarray}%
The ratio of growing to shrinking time is
\begin{equation}
\frac{\left\langle T^+\right\rangle }{\left\langle T^-\right\rangle }
= \alpha \left( \frac{1 + \frac{\mean{l}}{\alpha v \mean{T}}}
{1-\frac{\mean{l}}{v\mean{T}}}\right),
\end{equation}
where the r.h.s. is expected to stay close to the value $\alpha$,
as $l$ is smaller than $v\left\langle T^-\right\rangle$, except
for young rods. Because the number of growing time intervals is
close to the number of shrinking time intervals, there should not
be a large difference between the ratios $\left\langle
T^+\right\rangle/\left\langle T^-\right\rangle$ and $\left\langle
\tau^+\right\rangle/\left\langle \tau^-\right\rangle$, suggesting,
along with Eq.~\ref{eq:relationNgNsTaupTaum}, another relation:
\begin{equation}
\frac{\left\langle \tau^+\right\rangle }{\left\langle \tau^-\right\rangle }
\simeq
\frac{\left\langle T^+\right\rangle }{\left\langle T^-\right\rangle }
=\frac{n_g}{n_s } \simeq \alpha.
\label{eq:relationNgNsTaupTaumTpTm}
\end{equation}

To test this relation, we performed a set of simulations with the
same rate of injection $Q_i=1000$, and varying $\alpha$.  Both
$n_{g}$, $n_{s}$, $\left\langle \tau^{+}\right\rangle$,
$\left\langle \tau^{-}\right\rangle$, $\left\langle
T^{+}\right\rangle$ and $\left\langle T^{-}\right\rangle$ can be
independently obtained from the simulation, as summarized in
Figure~\ref{fig:populations}. The ratio $n_{g}/n_{s}$,
$T^{+}/T^{-}$, and $\left\langle
\tau^{+}\right\rangle/\left\langle \tau^{-}\right\rangle$ seems to
remain remarkably constant as the simulation goes on. The
agreement is not quantitative, but the four quantities of
relation~\ref{eq:relationNgNsTaupTaumTpTm} show significant
correlations~(Figure~\ref{fig:populations}).

\subsection{Anisotropic and non-stationary solutions of the kinetic theory}

There are indications that the stationary, homogeneous situation is not the
only possible solution of the kinetic theory. We outline in
Appendix~\ref{app:anisotropicSolution}, the main features of a
\textit{stationary} but \textit{anisotropic} solution, with a non trivial
dependence of $p_{gs}$ in the orientation~$\Omega$. This solution can explain
why the system tends to choose a preferential global orientation.  However,
the predicted anisotropy is less than the one that is numerically observed in
our system.

The kinetic model described above can also describe time dependent solutions.
However, we did not find any simple time-dependent solution compatible with our
boundary conditions, \textit{i.e.} a constant injection rate. It remains that
the existence of an unstable, non stationary and anisotropic solution of the
kinetic model cannot be ruled out.

\section{Conclusion}
\label{sec:conclusion}

We proposed and tested numerically a new paradigm for a kinetically
constrained growth mechanism of living rods. We demonstrated that the
collective behavior of the rods were leading to the formation of bundles of
well oriented rods, in the absence of excluded volume interactions and
chemical gradients. We suggest this mechanism as a possible alternative in the
formation and the orientation of microtubules gels.

We tested the responsive properties of the system to some external stress. We
found that the alignment was much faster in the presence of any small
anisotropic bias. This result is interesting in the context of the
experimental microgravity results of Tabony~\cite{TabonySim,TabonySim2}.
We developed a kinetic theory which accounts for the basic scaling properties
of the system: conversion rates, average length of the microtubules.  The
quantitative agreement remains poor, due to the strong correlations present in
this two-dimensional system, in which the emergence of large anisotropic
domains is incompatible with the homogeneity assumption ($\vec{r}$
independence).

Because we were able to show that this mechanism also induces some
alignment in three dimension, we believe that our kinetic theory
would give a better agreement in higher dimensional systems, which
is the subject of future work.


\appendix
\section{A kinetic theory for homogeneous distributions}
\label{app:kineticTheory}

We assume first that the system is homogeneous, and that the distribution
functions do not depend on $\vec{r}$.
Denoting by $S$ the area of the system, the distribution functions then reduce
to their homogeneous form:
\begin{equation}
c_{g,s,t}(l,\Omega, \vec{r},t) = \frac{1}{S} c_{g,s,t}(l,\Omega,t).
\end{equation}
The functions $c_{g,s,t}$ are ``extensive''functions of the area $S$.  We
propose a master equation for these functions, which accounts for the
shrinking and growing behavior of the rods, and also account for the
possibility of interconversion between the s and g states. In order to cope
with a possible global anisotropy of the rods, we let the interconversion
rates depend on the orientation $\Omega$, but not on the length:
$p_{gs}(\Omega)\dd t$ is the fraction of \textit{g}-rods which switches to a
\textit{s}-rod state during the time interval $\dd t$, while
$p_{sg}(\Omega)\dd t$ describes the reverse change. The corresponding master
equation reads:
\begin{equation}
\left\{
\begin{array}{l}
c_{g}(l,\Omega,t+dt)=c_{g}(l-vdt,\Omega,t)+\nonumber \\
p_{sg}(\Omega)c_{s}(l,\Omega,t)dt-
p_{gs}(\Omega)c_{g}(l,\Omega,t)dt; \\
c_{s}(l,\Omega,t+dt)=c_{s}(l+v^{-}dt,\Omega,t)+\nonumber \\
p_{gs}(\Omega)
c_{g}(l,\Omega,t)dt-p_{sg}(\Omega)c_{s}(l,\Omega,t)dt.
\end{array}
\right.  \label{eq:masterEquation2}
\end{equation}

The continuous  limit $\dd t\to 0$ leads to a system of two partial
differential equations:
\begin{equation}
\left\{
\begin{array}{c}
\displaystyle
\frac{\partial c_{g}}{\partial t}+v\frac{\partial c_{g}}{\partial l}%
=p_{sg}c_{s}-p_{gs}c_{g}; \\
\displaystyle
\frac{\partial c_{s}}{\partial t}-\alpha v\frac{\partial c_{s}}{\partial l}%
=p_{gs}c_{g}-p_{sg}c_{s}.
\end{array}
\right.  \label{eq:masterEquation3}
\end{equation}
where appears the ratio $\alpha=v^{-}/v=v^{-}/(v^+-v^-)$.

The partial derivatives in the left hand sides of
Eq.~\ref{eq:masterEquation3}, correspond to a drift motion of the
rods along the $l$ axis. We can associate to this drift the
``currents'' of growing $j_g(l,\Omega,t)=v c_g$ and shrinking
$j_s(l,\Omega,t)=-\alpha v c_s$ rods. The sum $j_g + j_s$ measures
the difference between the number of rods which have grown bigger
than $l$, and the number of rods which have shrunk below $l$. The
distribution $c_t=c_g+c_s$ obeys a usual conservation equation
$\partial c_t/\partial t + \partial (j_g+j_s)/\partial l=0$,
provided $l>0$.

In particular, as there is no other option for a rod with length 0 than
growing or disappearing, and our master equations must be completed with a
boundary condition involving $j_g$ and the injection rate $Q_i$.
\begin{equation}
Q_i(\Omega) = v c_g(l=0,\Omega).
\label{eq:boundaryQg}
\end{equation}
Meanwhile, the rate of disappearance of the rods is $Q_d$, obeying
\begin{equation}
Q_d(\Omega) = \alpha v c_s(l=0,\Omega).
\label{eq:boundaryQs}
\end{equation}
At the other extreme, we expect that
\begin{equation}
\lim_{l\to \infty} c_{g,s,t}(l,\Omega)= 0.
\label{eq:boundary:infty}
\end{equation}
Equations
\ref{eq:masterEquation2},\ref{eq:boundaryQg},\ref{eq:boundary:infty}
are the basis of our kinetic theory.

\section{The stationary case}
\label{app:stationaryCase}

The above system of equations is simpler if we look for a time independent
solution, setting the time partial derivative to zero. Introducing
$c_t=c_g+c_s$, and $c_d=c_g-c_s$, we get:

\begin{equation}
\left\{
\begin{array}{l}
\displaystyle
\frac{v}{2}(1-\alpha )\frac{\partial c_t}{\partial l}+\frac{v}{2}(1+\alpha )%
\frac{\partial c_d}{\partial l}=0; \\
\displaystyle
\frac{v}{2}(1-\alpha )\frac{\partial c_d}{\partial l}+\frac{v}{2}(1+\alpha )%
\frac{\partial c_t}{\partial l}= \nonumber \\
(p_{sg}-p_{gs})c_t-(p_{sg}+p_{gs})c_d; \\
\displaystyle c(0,\Omega)=q;\; \lim_{l\to\infty} c(l,\Omega)=0,
\end{array}
\right.
\end{equation}
with solution
\begin{equation}
\left\{
\begin{array}{l}
c_t(l,\Omega )=qe^{-l/\overline{l}(\Omega )}; \\
c_d(l,\Omega )=\frac{\alpha -1}{1+\alpha }qe^{-l/\overline{l}(\Omega )},%
\end{array}%
\right.  \label{cl}
\end{equation}
where, for convenience, we have introduced  $q=c_t(0,\Omega)$,
while the average length in the direction $\Omega $ is given by:
\begin{equation}
\overline{l}(\Omega )=\frac{\alpha v}{\alpha p_{gs}(\Omega)-p_{sg}(\Omega)}.%
\label{eq:lrel}
\end{equation}
Moreover, the distributions $c_g$ and $c_s$ verify
\begin{equation}
\frac{c_{g}(l,\Omega )}{c_{s}(l,\Omega )}=\alpha,  \label{cacd}
\end{equation}
and consequently,
\begin{eqnarray}
c_g &=& \frac{\alpha q}{\alpha+1} e^{-l/\overline{l}(\Omega )};\nonumber\\
c_s &=& \frac{q}{\alpha+1} e^{-l/\overline{l}(\Omega )}.
\end{eqnarray}
Finally, the injection rate $Q_i(\Omega)$  is related to $q(\Omega)$ by
\begin{equation}
  Q_i(\Omega) = \frac{\alpha}{\alpha+1} v q(\Omega),
\end{equation}
which is the basis of a stationary, homogeneous solution of the
system, expressed in terms of $Q_i$, $\alpha$, $p_{gs}$ and
$p_{sg}$.  An example of explicit angular dependence of
$Q_i(\Omega)$ is the situation described in
section~\ref{subsec:sensitivity} and in
Figure~\ref{fig:effectBias}. In most cases, however, we are
interested in an isotropic, constant, function $Q_i$, which we
consider now.

To move further, we must estimate the interconversion rates
$p_{gs}$ and $p_{sg}$. To estimate $p_{sg}$, we assume that
collisions are pairwise, and, as in the usual kinetic theory of
gases, that there is no correlation between any two colliding
rods. Then, $p_{sg}$ does not depend on angles, and is inversely
proportional to the average waiting time $\langle\tau^-\rangle$
spent in the blocked state.  Since all rods shrink at the same
speed, the waiting time depends only on the distance between the
contact point and the minus end of the restricting rod. Assuming a
uniform distribution of contact points along the rod, we find that
the average waiting time associated to a restricting rod with
length $l$ is $l/(2v^-)$, and consequently,

\begin{equation}
\mean{\tau^{-}} = \frac{\overline{l}}{2v^{-}};\;
p_{sg}=\frac{1}{\mean{\tau^{-}}}=\frac{2v^{-}}{\overline{l}}.%
\label{eq:psg}
\end{equation}
In this equation, we need to know the average length $\overline{l}$ of the
rods, irrespective of their orientation. Given the solution obtained above,
this simply reads:
\begin{equation}
\overline{l}=\frac{\int\dd l\dd\Omega\, l\, c_t(l,\Omega )}
{\int\dd l\dd\Omega\, c_t(l,\Omega )}
=\frac{\int\dd\Omega \Big(\overline{l}(\Omega)\Big)^2}
 {\int\dd\Omega\,\overline{l}(\Omega )}. \label{llOmega}
\end{equation}

The estimate of $p_{gs}$ also comes from the analogy with the
kinetic theory of gases. We imagine the system from the point of
view of an observer sitting at the top of a growing tip, and
estimate the area swept by the mesh of all the other rods, moving
relatively to the observer at speed $-v^+$. The typical collision
time is reached when this area becomes comparable to the total
area $S$  of the system, making the probability of collision of
order one. The calculation shows that the collision time depends
on the projected length $l^{\prime}$ of the obstructing rod, and
on the relative orientation difference $\Omega -\Omega ^{\prime }$
between the two rods.
We can write:
\begin{equation}
\overline{l_{p}}(\Omega )=\frac{\int d\Omega ^{\prime }dl^{\prime
}l^{\prime} \left\vert \sin (\Omega -\Omega ^{\prime })\right\vert
c_t(l^{\prime },\Omega ^{\prime })}{\int d\Omega ^{\prime }dl^{\prime
}c_t(l^{\prime },\Omega ^{\prime })},  \label{eq:lp}
\end{equation}
and the probability $p_{gs}(\Omega )$ reduces to:
\begin{equation}
p_{gs}(\Omega )=v^{+}\overline{l_{p}}(\Omega )\left(
\frac{n_t}{S}\right), \label{eq:pgs}
\end{equation}%
where $S$ is the total area of the system, $n_t$ is the total number
of rods, and $(n_t/S)$ is the ratio of two ``extensive'' functions.

\section{The isotropic solution and its predictions}

The isotropic hypothesis consists in taking $p_{gs}$ and $p_{sg}$ angle
independent. This simplifies the above kinetic theory to a point where a
self-consistent analytical solution becomes available. The $\Omega$ dependence
of $\overline{l}$ and $\overline{l}_p$ drops out, and we get:
\begin{equation}
\overline{l}_p =\frac{2\overline{l}}{\pi},
\label{eq:lp-isotropic}
\end{equation}
and equation~\ref{eq:lrel} leads to the self-consistence
relations:

\begin{eqnarray}
\overline{l} &=&\frac{ \alpha v}{\alpha v^+ \displaystyle\frac{n_t}{S}
\frac{2\overline{l}}{\pi} -\frac{2v^-}{\overline{l}}}; \\
\Big(\overline{l}\Big)^2 &=& \frac{3\pi}{2(\alpha+1)}\frac{S}{n_t}.
\label{eq:isotropic-self-consistence}
\end{eqnarray}

Then, we replace $n_t$ by $\frac{\alpha+1}{\alpha}\frac{Q_i\overline{l}}{v}$,
to make a prediction for the average length $\overline{l}$ and the number of
rods~$n_t$.

\begin{eqnarray}
\overline{l} &=& \sqrt{\frac{3\pi}{2(\alpha+1)}\frac{S}{n_t}} \simeq
\left( \frac{S}{n_t}\right)^{\frac{1}{2}};\\
\overline{l} &=& \sqrt[3]{\frac{3\alpha}{4(\alpha+1)^2}\frac{Sv}{Q_i}}
\simeq \left( \frac{S v}{Q_i}\right)^{\frac{1}{3}};\\
n_t &=& 2\pi\frac{\alpha +1}{\alpha v}\, \overline{l}\, Q_i.
\end{eqnarray}
%

\section{The anisotropic solution}
\label{app:anisotropicSolution}

We believe that the system of equations~\ref{eq:lrel},
\ref{eq:psg}, \ref{llOmega}, \ref{eq:lp}, \ref{eq:pgs} also admits
an anisotropic solution, characterized by an explicit $\Omega$
dependence of $p_{gs}(\Omega)$ and $\overline{l}(\Omega)$ while
$p_{sg}$ remains isotropic. To approach this solution, we expand
$\overline{l}(\Omega)$ in cosine series:
\begin{equation}
\overline{l}(\Omega) = l_0 + l_2 \cos(2\Omega) + l_4 \cos(4\Omega)\ldots
\end{equation}
and approximate $\sin|\Omega-\Omega'|$ in a similar manner:
\begin{equation}
\sin|\Omega| = s_0 + s_2 \cos(2\Omega) + s_4 \cos(4\Omega)\ldots
\end{equation}
Possible choices include the Fourier expansion:
\begin{equation}
\sin|\Omega| = \frac{2}{\pi}
-\frac{4}{\pi}\sum_{p=1}^{\infty}\frac{\cos(2p\Omega)}{4p^2-1},
\end{equation}
or replacing $\sin|\Omega|$ by $\sin^2(\Omega)$.

For instance, by keeping the two first terms in the expansion, and
writing the self-consistence equation \ref{eq:lrel} under the form
$\overline{l}(\Omega )\times (\alpha p_{gs}(\Omega)-p_{sg})=\alpha
v$, we finally obtain a system of equations for $l_0$ and
$x=l_2/l_0$:
\begin{eqnarray}
\frac{\alpha v^+ n_t l_0^2}{S} \left(
  s_0\left[1+\frac{x^2}{2}\right]+s_2\frac{x^2}{2} \right) &=& 3v^-;\nonumber\\
\frac{\alpha v^+ n_t l_0^2}{S} \left( s_0 x\left[1+\frac{x^2}{2}\right] +s_2
  x\right) &=& \frac{2 v^- x}{1+\frac{x^2}{2}}.
\end{eqnarray}

We observe that the isotropic solution $x=0$, $l_0^2= 3Sv^-/(s_0\alpha v^+
n_t)$ (equivalent to \ref{eq:isotropic-self-consistence})  coexists along
with an anisotropic solution $x\neq 0$, where $x$ solves
\begin{equation}
\frac{\alpha v^+ n_t l_0^2}{S}
\left( s_0\left[1+\frac{x^2}{2}\right] +s_2\right) = \frac{2
  v^-}{1+\frac{x^2}{2}},
\end{equation}
and $l_0$ is a function of $x$ and the other parameters of the problem.

Thus, despite its strong mean field features, the kinetic model is compatible
with the emergence of an anisotropic solution, with an explicit angular
dependence of the average length of the rods. However, this anisotropy is
bounded, with an average length finite in all directions, while the domains
observed in the simulations can grow without limit.

\begin{acknowledgements}

The authors wish to thank Professor J. Tabony for discussions
inspired this work. V. B. gratefully acknowledges Centre National
d'Etudes Spatiales (CNES) for a research post-doctoral fellowship.

\end{acknowledgements}


\end{document}